\documentclass[twocolumn,aps,prl,superscriptaddress, amsmath,amssymb]{revtex4-2}
\usepackage{graphicx}
\usepackage{dcolumn}
\usepackage{bm}
\usepackage{epstopdf}
\usepackage{xcolor}
\usepackage{siunitx} 
\usepackage[normalem]{ulem}
\usepackage{lineno}
\usepackage{amsmath} 

\usepackage{upgreek}  
\begin{document}
\title{Synthetic active liquid crystals powered by acoustic waves}

\author{Andrey Sokolov}
\email{sokolov@anl.gov}
\affiliation{Materials Science Division, Argonne National Laboratory, 9700 South Cass Avenue, Lemont, IL 60439, USA}

\author{Jaideep Katuri}
\affiliation{Materials Science Division, Argonne National Laboratory, 9700 South Cass Avenue, Lemont, IL 60439, USA}

\author{Juan J. de Pablo}
\affiliation{Materials Science Division, Argonne National Laboratory, 9700 South Cass Avenue, Lemont, IL 60439, USA}
\affiliation{Pritzker School of Molecular Engineering, University of Chicago, Chicago, IL 60637, USA}

\author{Alexey Snezhko}
\affiliation{Materials Science Division, Argonne National Laboratory, 9700 South Cass Avenue, Lemont, IL 60439, USA}

\begin{abstract}

Active nematics are materials composed of mobile, elongated particles that can transform energy from the environment into a mechanical motion. Current experimental realizations of the active nematics are of biological origin and include cell layers, suspensions of elongated bacteria in liquid crystal, and combinations of bio-filaments with molecular motors \cite{sanchez2012spontaneous,zhou2014living}. Here, we report the realization of a fully synthetic active nematic system comprised of a lyotropic chromonic liquid crystal energized by ultrasonic waves. This artificial active liquid crystal is free from biological degradation and variability, exhibits stable material properties, and enables precise and rapid activity control over a significantly extended range. We demonstrate that the energy of the acoustic field is converted into microscopic extensile stresses disrupting long-range nematic order and giving rise to an undulation instability, development of active turbulence, and proliferation of topological defects. We reveal the emergence of unconventional free-standing persistent vortices in the nematic director field at high activity levels.  The results provide a foundation for the design of externally energized active nematic fluids with stable material properties and tunable topological defects dynamics crucial for the realization of reconfigurable microfluidic systems.

\end{abstract}

\maketitle


\begin{figure*}[t]
\centering
\includegraphics[width = 1.0\textwidth]{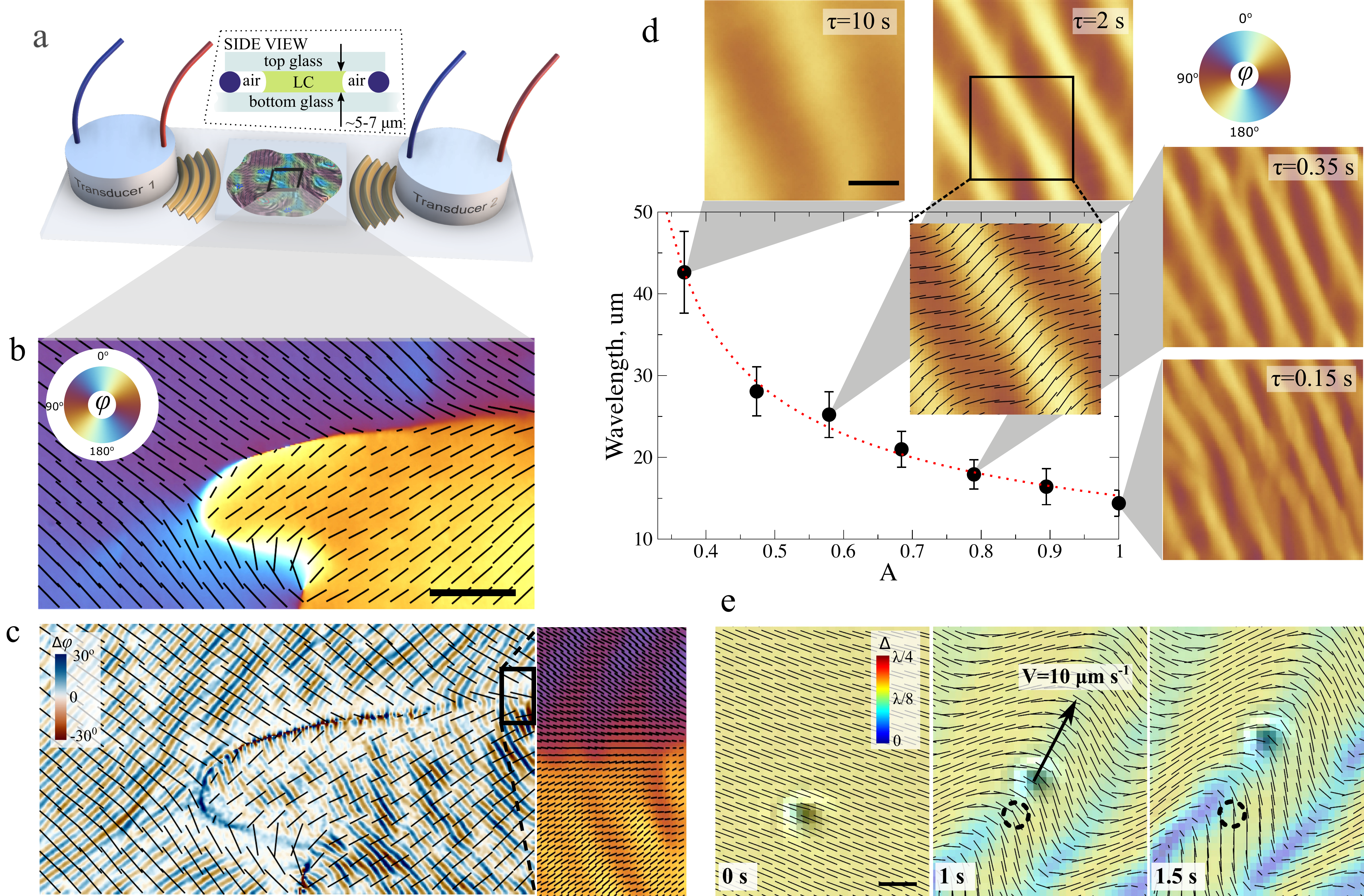}
\caption{\label{fig1} \textbf{Acoustically energized liquid crystals.} (a) Schematic representation of the experimental setup. A droplet of DSCG is sandwiched between two glass slides and acoustically energized by two transducers attached to the bottom slide. (b) An example of the spatial configuration of the director (black line segments) in two distinct domains before the ultrasound was applied. Color depicts the local orientational angle $\varphi$ according to the color-map in the top left corner. The scale bar is 50 $\mu$m. (c) The deflection angle $\Delta \varphi$ of the local director upon application of the excitation signal. The stripes of the undulations are normal to the local orientation of the director. (d) Dependence of the wavelength of the observed undulation on the amplitude of the acoustic field. The dot line shows the fit $\sim 1/ \sqrt{A-a_0})$ where $a_0=0.27$. The inset images demonstrate patterns that emerged in DSCG at the corresponding amplitudes of the acoustic field. The scale bar is 20 $\mu$m. The error bars correspond to standard deviations of the mean values.
(e) The advection of a microscopic particle by the flow produced by undulation instability upon application of the acoustic field. The diameter of the particle is 3 $\mu$m. The color indicates the local optical retardance. The scale bar is 5 $\mu$m.
}
\end{figure*}

The diverse phenomenology of active nematics arises from a combination of the anisotropic characteristics of liquid crystals and their non-equilibrium behavior. The hallmark feature of active nematics is a continuous reconfiguration of the orientational order and the spatiotemporal variations of the order parameter, frequently called 'active turbulence', arising from the interplay between anisotropic viscoelastic forces and generated active stresses. 
Active turbulence in nematic materials inevitably coexists with another intriguing phenomenon - continuous proliferation and annihilation of topological defects \cite{thampi2014instabilities,giomi2014defect,kumar2022catapulting}, localized singularities of the order parameter, and is a subject of extensive study.
 The existing experimental models of active nematics are either fully biological systems or bio-synthetic composites \cite{doostmohammadi2018active,zhang2021autonomous}. One example is a mixture of filamentary proteins and molecular motors that generate activity through adenosine triphosphate (ATP) hydrolysis converting chemical energy into a mechanical motion \cite{ndlec1997self,sanchez2012spontaneous}. Another system, named Living Liquid Crystal (LLC), is a composite of a water-based liquid crystal and live swimming bacteria \cite{zhou2014living}. The hydrodynamic flows created by bacteria flagella exert local microscopic stress on the liquid crystal, causing disturbances in a long-range nematic order. The activity level in these two systems is controlled by the ATP and bacterial concentration, respectively. Both of these biological active nematic systems exhibit inherent variability due to differences in biological components and are subject to unavoidable biological aging (degradation). Moreover, bio-synthetic systems have a very limited range of accessible activity levels \cite{zhang2021spatiotemporal,ruijgrok2021optical}. 
 An artificial active nematic system  with well-controlled and stable material properties would be able to resolve both these deficiencies.

Here, we report the realization of a fully synthetic active nematic system  - acoustically energized liquid crystal (AELC) - free from biological agents or supplementary motile components, and with the activity level controlled externally in a wide range by the amplitude of the acoustic field.
We experimentally demonstrate the conversion of the acoustic energy into local extensile stresses of the liquid crystalline media, resulting in the emergence of undulation instability with the characteristic wavelengths governed by the amplitude of the acoustic field. As the activity increases, the system transitions into a turbulent-like state characterized by a chaotic generation, motion, and annihilation of topological defects. We reveal the spontaneous formation of the unconventional free-standing persistent hydrodynamic vortices at high activity levels not previously observed in biologically-based active nematic systems.

\begin{figure*}[t]
\centering
\includegraphics[width = 1.0\textwidth]{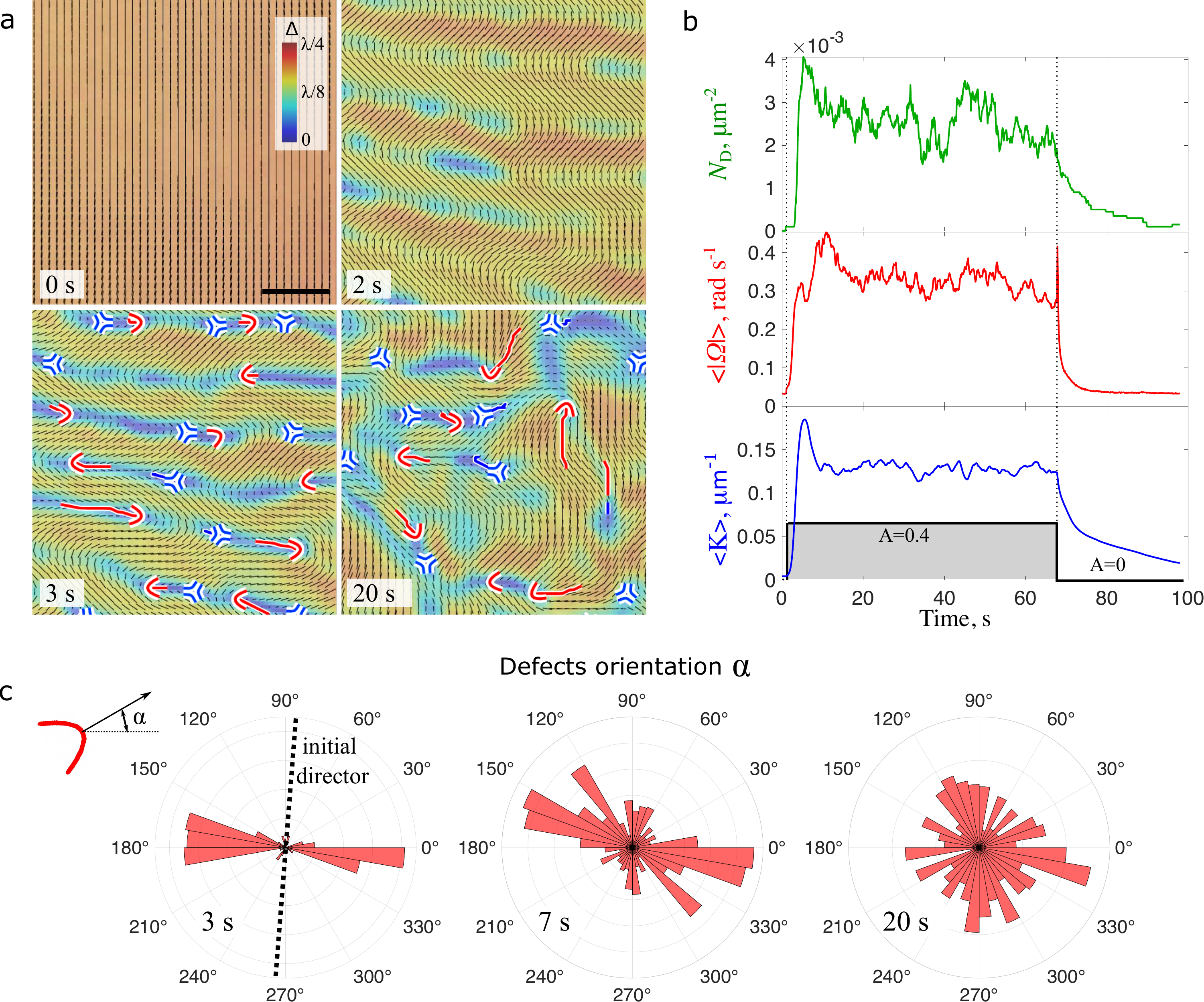}
\caption{\label{fig2} \textbf{Proliferation of topological defects and the emergence of active turbulence in acoustically powered liquid crystal.} (a) Reconstructed director fields (short black lines) and the optical retardance $\Delta$ (color) at different moments elapsed since the application of the acoustic field. Topological defects are shown by red (+1/2 defect) and blue (-1/2 defect) symbols. The trajectories of positive and negative defects over one second are indicated by red and blue lines, respectively. The scale bar is 15 $\mu$m. (b) Temporal evolution of the average director rotation rate $\langle | \Omega | \rangle$, curvature $K$ and defects density $N_D$ upon application of the acoustic field. The AELC is energized between 1 s and 68 s at the amplitude of $A=0.4$. Non-zero $\Omega$ before and after the active period reflects the thermal fluctuations of liquid crystal molecules and camera noise. 
(c) A polar histogram of positive topological defect orientations shown in (a) at different times (3 s, 7 s, and 20 s) elapsed since the application of the acoustic field. Initially, the defects are aligned with the emerging undulations resulting in anisotropic distribution of the defect orientations (left panel). As the system evolves toward the active turbulence state the distribution becomes isotropic.  
}
\end{figure*}

The experimental system consists of the lyotropic chromonic liquid crystals DSCG (disodium cromoglycate) sandwiched between two glass plates with a pair of ultrasonic transducers glued to the bottom glass, see Fig.\ref{fig1}a. For the sufficiently thin films ( thickness is less than 20 $\mu$m), the orientation of the director remains in the plane of the experimental glass plates \cite{zhou2014living,park2012lyotropic} and can be reconstructed from the birefringence map captured by a digital polarization camera using 2D approximation. An example of a reconstructed director field is shown in Fig. \ref{fig1}b, where two distinctive domains of the director field are observed. When the transducers are energized by a sinusoidal signal at ultrasonic frequencies (126 kHz for results shown in Fig. \ref{fig1}c), the acoustic waves propagate through the glass and the LC medium, inducing re-configurations in the initially uniform director field in each domain, see also  Supplementary Movie 1.
Within a fraction of a second, we observed a formation of a stripe pattern formed by periodic director undulations (Fig. \ref{fig1}c,d), similar to those caused by local extensile stresses in other nematic active systems such as Living Liquid Crystal \cite{zhou2014living}, and mixtures of bio-filaments and molecular motors \cite{gao2015multiscale,martinez2019selection,senoussi2019tunable,simha2002hydrodynamic,voituriez2005spontaneous}.

The bending instability in active nematic systems is usually accompanied by induced hydrodynamic flows normal to the local orientation of the director field ~\cite{gao2015multiscale,martinez2019selection,zhou2014living}. To confirm the presence of similar flows in our system, we added small tracers (2.2 $\mu$m in diameter) and tracked their motion upon application of the acoustic field, see Fig. \ref{fig1}e and Supplementary Movies 2 and 3. The direction of the tracers' motion matched the director field bending.

Experiments performed at various amplitudes and frequencies of the excitation signal reveal two key features of the observed bending instability: i) the orientation of the stripes is perpendicular to the initial local director field, as shown in Fig. \ref{fig1}(b,c), and ii) the wavelength ($\lambda$) of the bending instability decreases with the amplitude of the applied voltage ($A$, normalized by maximum voltage of 20V) as $\lambda \sim 1/\sqrt{A-a_0}$, where $a_0$ is a small constant (Fig. \ref{fig1}d) corresponds to a minimum amplitude of the acoustic field resulting in the emergence of undulations in our experimental cell.
Such response of the system qualitatively and quantitatively mimics the behavior of active nematic systems with extensile stress, such as Living Liquid Crystals or microtubules-molecular motors systems \cite{simha2002hydrodynamic,ramaswamy2007active,doostmohammadi2018active,zhou2014living}.

Analogous to biological active nematic systems, the injection of activity in the liquid crystal by the acoustic field leads to the spontaneous topological defect proliferation causing the destruction of long-range nematic order and the emergence of an "active turbulence" state characterized by a chaotic-like behavior of topological defects, see Fig.~\ref{fig2}a and supplementary Movie 4. 
The mechanism of bending instability is well-understood in biological active nematic systems and is the result of active extensile stress $\mathbf{\Sigma}^{a}=\alpha \mathbf{Q}$, where $\alpha$ is activity parameter ( $\alpha<0$ corresponds to an extensile stress) and $\mathbf{Q}=S(n_j n_j-\delta_{ij}/2)$ is an orientational tensor describing local alignment of liquid crystal molecules $\mathbf{n}$ and the local order parameter $S$ \cite{simha2002hydrodynamic,thampi2016active,hemingway2016correlation,srivastava2016negative}. In these systems the active stress $\mathbf{\Sigma}^a$ is generated by hydrodynamic flows created by swimming bacteria or forces exerted by molecular motors with the activity parameter $\alpha$ proportional to the magnitudes of these flows or forces. The observations of the acoustically powered LC system suggest that the ultrasonic wave in our experimental setup generates similar extensile microscopic stresses $\mathbf{\Sigma}^a$ that drive the liquid crystal out of equilibrium, and the absolute value of the activity parameter $|\alpha|$ increases linearly with the applied voltage to the ultrasonic transducer.

\begin{figure*}[t]
\centering
\includegraphics[width = 1\textwidth]{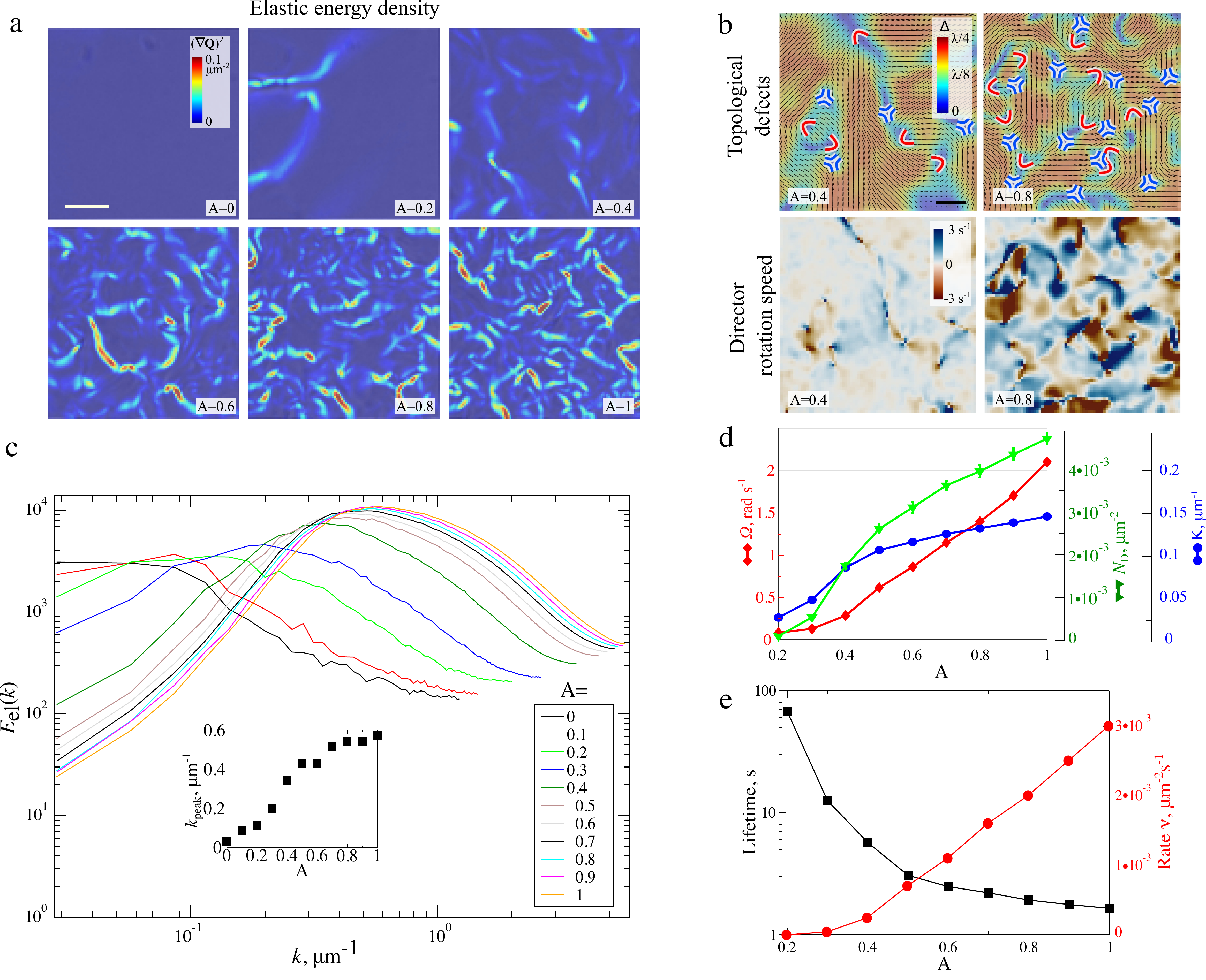}
\caption{\label{fig3} \textbf{Spatiotemporal properties of acoustically energized liquid crystal in an active turbulent state.} (a) Spatial distribution of the elastic energy density for various activity levels. (b) Top:  Director field (black lines) and phase retardance maps for A=0.4 and A=0.8. Bottom: Distribution of the local director rotational speed. (c) Timed averaged elastic energy spectrum at different activity levels. Inset: Position of the elastic energy spectrum maximum as a function of activity. (d) Dependence of the space- and time-averaged magnitude of the rotation speed $\Omega$, the local curvature $K$ of the director field, and the topological defects density $N_D$ on the activity $A$. (e) The average lifetime of defects (black squares) and the defect generation/annihilation rate (red circles) at different amplitudes of the acoustic field. 
}
\end{figure*}

\begin{figure*}[t]
\centering
\includegraphics[width = 1.0\textwidth]{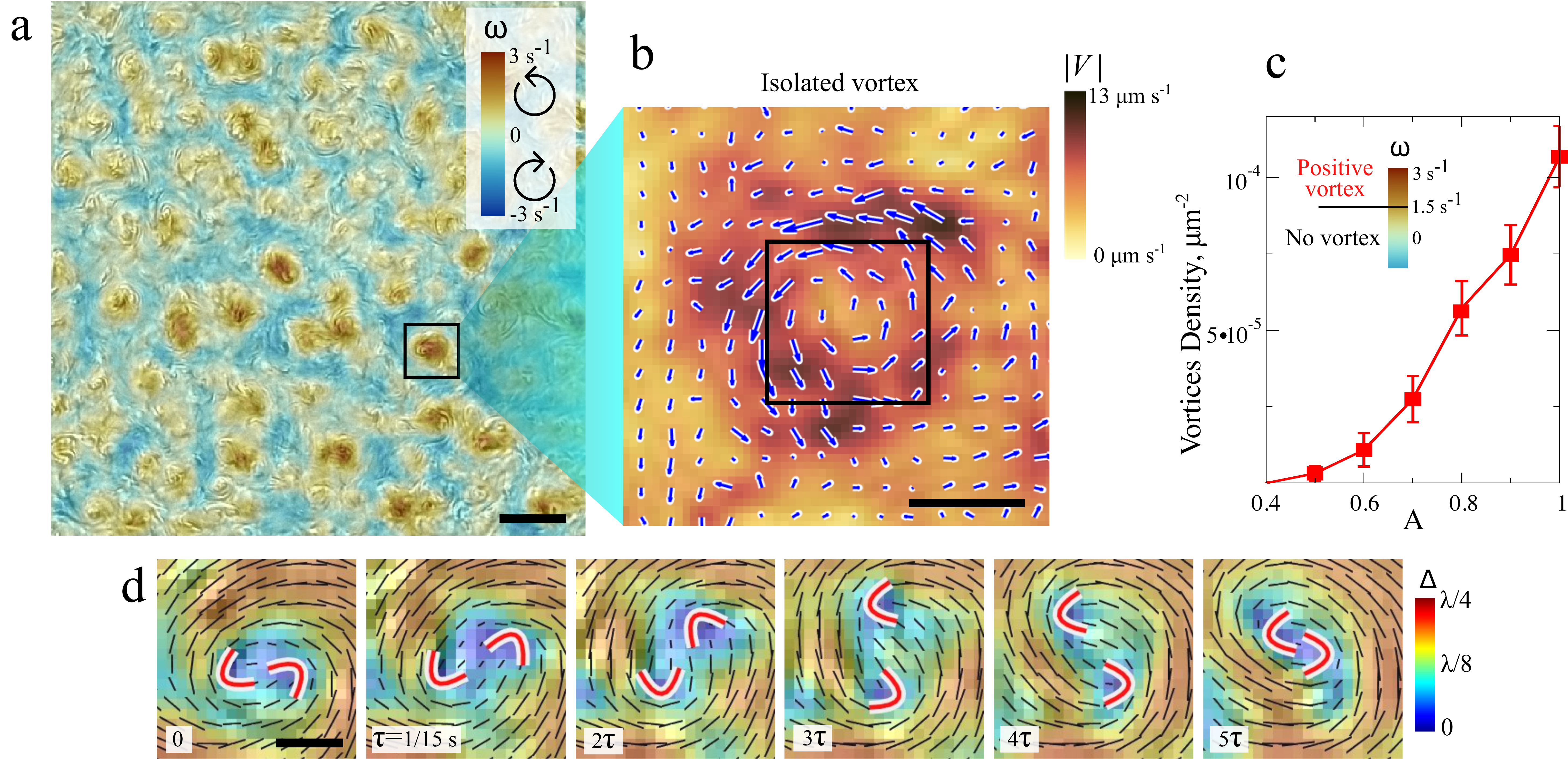}
\caption{\label{fig4} \textbf{Free-standing vortices.}  (a) a hydrodynamic vorticity map superimposed over the original raw image at $A=0.8$. Brown and blue colors correspond to CCW and CW rotations respectively. The scale bar is 50 $\mu m$. See also supplementary movie 5. (b) The velocity field obtained by PIV in the vicinity of a single isolated vortex. The background color depicts the magnitude of the local velocity. The scale bar is 10 $\mu m$ (c) The number of vortices in the field of view shown in (a) at different activity levels $A$. The vortex is defined as an isolated area with the peak local vorticity above 1.5 $s^{-1}$. The error bars correspond to standard deviations of the mean. (d) Rotation of two positive defects in the core of the vortex. See also supplementary movie 6. The scale bar is 5 $\mu m$. 
}
\end{figure*}

While the evidence of extensile active stresses in the AELC system is well-documented, the mechanisms responsible for the conversion of acoustic energy into a mechanical motion of the LC director field are not obvious. In general, ultrasound waves affect liquid crystals in various ways, such as inducing flow, altering the orientation of the liquid crystal molecules, and creating static patterns in the liquid crystal structure. For example, it was previously shown that the orientation of LC molecules can be changed statically and uniformly in space by acoustic radiation force \cite{koyama2012ultrasonic,shimizu2018ultrasound,liu2013holographically,selinger2002acoustic,witkowska1983optically}.
Reorientation of LC molecules \cite{ozaki2007reorientation} and the emergence of periodic stripes and patterns were also attributed to acoustic streaming as a result of ultrasonic compression field \cite{kapustina2004liquid,kapustina2008ultrasound,sripaipan1977ultrasonically}. In all these works, the period of the observed static patterns is defined either by the thickness of the experimental cell \cite{kapustina2004liquid,kapustina2008ultrasound} or the excitation frequency \cite{sripaipan1977ultrasonically}. In addition, the orientation of the observed stripes is parallel to LC/air interface.
Periodical domain structures in LC (5CB) may also be induced by the elastic wave propagating in the glass plate \cite{moritake1999acoustooptic,ozaki2006evaluation}. However, the wavelength decreases with the excitation frequency (2-25MHz range) and is observed either parallel or perpendicular to the acoustic wave direction. 

None of these previously reported phenomena apply to the observed behavior. In our system, we did not detect any averaged flow preceding the observed undulation instability. The stripes emerge always perpendicular to the local orientation of the director field, and what is more important, the period of the striped structure is controlled by the amplitude of the acoustic wave, not the frequency or geometry of the experimental setup. In addition, only lyotropic chromonic liquids such as DSCG or Sunset Yellow showed the bending instability and other phenomena described below, while other liquid crystals like MBBA, 5CB, and 8CB did not exhibit these features. These experimental observations suggest that the possible mechanism for the generation of active stresses may be related to the tumbling behavior of DSCG \cite{baza2020shear} enabled by high-frequency streaming flows induced by the vibrating plates of the cell.

As the level of activity increases with the amplitude of the acoustic wave, the undulations transition into an active turbulent state, see Fig. \ref{fig2}a and supplementary Movie 4. In active nematic systems the turbulent state is manifested by a chaotic reconfiguration of the director field and continuous proliferation/annihilation of topological defects \cite{schaller2013topological,keber2014topology,sanchez2012spontaneous,giomi2014defect,bowick2022symmetry,giomi2015geometry}. The rich phenomenology of active turbulence comes from the complexity of active nematodynamics \cite{de1993physics,peng2016command,mushenheim2014dynamic,bade2018edges}.  
Because of the distinct geometrical structure of topological defects, the initial positive defects align parallel to the bending stripes and perpendicular to the original orientation of the director field, see Fig.~\ref{fig2}c. The positive defects start to move along the stripes while negative defects remain mostly stationary. Eventually, the system reaches a well-developed active turbulence state, and the orientation of positive defects becomes isotropic, Fig. \ref{fig2}(a,c) and independent of the initial orientation of the director field. Once the activity is removed, the active turbulent state relaxes back to a uniformly aligned configuration. 
To quantify the evolution of the nematic order upon application of the acoustic field we calculate the angular speed $\Omega=\langle|\dot{\mathbf{n}}|\rangle$ and the curvature $K=\langle |(\mathbf{n} \cdot \nabla) \mathbf{n} | \rangle$ of the director field averaged over the observation area, as well as the density of topological defects $N_D$. A typical temporal evolution of these parameters in AELC is shown in Fig. \ref{fig2}b. The system reaches a steady dynamic state at the order of 10s. 
Interestingly,  $\Omega$ exhibits a distinct peak when the system is de-energized which is also observed upon energizing the system. The spike is linked to a rapid relaxation of the elastic stresses in areas corresponding to a non-uniform structure of the director field (such as topological defects) upon removal of the external excitation. The director field undergoes a local rapid reconfiguration in the areas with the large elastic stress. Naturally, a strong spatial correlation between the local elastic energy and the speed of the director rotations is expected and observed, see Fig.~\ref{fig3}a,b. 
The ability to control activity on demand allows us to probe the active turbulence states at a wide range of activity levels in a single experiment (often not achievable in bio-based systems). Characteristic reconfiguration speed of the director field $\Omega$, the average curvature $K$, and the defects density $N_D$ in a steady active turbulent state as a function of the activity $A$ are presented in Fig \ref{fig3}b,d. The linear increase of $\Omega$ indicates a direct proportionality between the rotational speed of the local director field and the hydrodynamic vorticity, which grows linearly with the activity level \cite{giomi2015geometry}.
Furthermore, the number of topological defects $N_D$ demonstrates almost linear increase with the activity, consistent with previous theoretical findings \cite{giomi2015geometry,shankar2019hydrodynamics}. The defect generation rate $\nu$  increases with the activity, while the average lifetime of the defects decreases, see Fig. \ref{fig3}e, due to an increase in the defect recombination probability.

We further investigated an elastic energy spectrum and its relationship to the activity by extracting the spatial distribution of elastic energy $(\nabla Q)^2$ in a 2D film of AELC, see Fig.~\ref{fig3}a. A typical spectrum has a peak that shifts towards smaller scales with the activity (see Fig.\ref{fig3}c) in agreement with theoretical predictions for nematic systems~\cite{krajnik2020spectral,rorai2022coexistence}.


The analysis of hydrodynamic vorticity in acoustically powered liquid crystal reveals an unexpected phenomenon - the emergence of free-standing vortices at the high activity levels ($A>0.5$), see Fig.\ref{fig4}a,b and supplementary Movies 5,6 and 7. Such vortices appear as localized round regions exhibiting high vorticity values. 
A typical vortex size is in the range from 10 $\mu m$ to 25 $\mu m$ (the vorticity magnitude threshold of 1.5 $s^{-1}$ was set to define the vortex size). 
Surprisingly, the majority of vortices within the same active domain exhibit consistent rotation in a single direction (clockwise or counterclockwise), nevertheless, no overall domain rotation is observed (the overall vorticity of a domain is zero) and the polarity of vortices can alternate between different domains, see Supplementary Movie 10. The direction of a vortex rotation in domains can also be reversed by changing the acoustic field frequency (see Supplementary Movie 11). The intricate mechanism behind the spontaneous vortex generation and sorting into domains is not clear at this point and presents an exciting avenue for future research.
The vortex spatial density increases with the activity $A$, Fig. \ref{fig4}c. The core of a typical vortex contains two circulating $+1/2$ defects (see Fig. \ref{fig4}d and supplementary Movie 6), resembling the observations in constrained active nematic systems \cite{opathalage2019self}.

In summary, we introduce a fully synthetic active nematic system comprised of a thin film of lyotropic chromonic liquid crystal (DSCG) energized by an ultrasound wave. The system is free from biological limitations (degradation, aging, actuation through consumption of a chemical fuel, limited range of activities) and demonstrates all phenomenology associated with active nematics - undulation instability,  generation of topological defects, and active turbulent motion. Acoustically powered liquid crystals have stable physical properties and enable precise and rapid activity control over a significantly extended range. Access to significantly higher activity levels revealed the emergence of clusters of coherent nematic vortices. The results provide a foundation for the unorthodox design of artificial active nematic fluids with {stable and fully controlled material properties}.

\section*{Acknowledgements}
The research was supported by the U.S. Department of Energy, Office of Science, Basic Energy Sciences, Materials Sciences and Engineering Division. 

\section*{}
The submitted manuscript has been created by UChicago Argonne, LLC, Operator of Argonne National Laboratory (“Argonne”). Argonne, a U.S. Department of Energy Office of Science laboratory, is operated under Contract No. DE-AC02-06CH11357. The U.S. Government retains for itself, and others acting on its behalf, a paid-up nonexclusive, irrevocable worldwide license in said article to reproduce, prepare derivative works, distribute copies to the public, and perform publicly and display publicly, by or on behalf of the Government. The Department of Energy will provide public access to these results of federally sponsored research in accordance with the DOE Public Access Plan. http://energy.gov/downloads/doe-public-access-plan

\bibliography{refs}
\bibliographystyle{rsc}

\end{document}


\title{Supplemental Material for\\ ``Acoustically energized active liquid crystal''}

\maketitle

\makeatletter 
\renewcommand{\thefigure}{S\@arabic\c@figure}
\makeatother

\section{Methods}
\textbf{Sample preparation.}
The experimental system consists of a small droplet (0.2 $\mu$L) of a lyotropic chromonic liquid crystal, disodium cromoglycate (DSCG, 13\% by weight) water dispersion, confined between two glass slides and exposed to ultrasonic waves, see Fig \ref{fig1}a. Two ultrasonic transducers (H2KLPY11000600, DigiKey) are attached to the bottom slide and driven by a function generator (Agilent 33210a).  The LC droplet thickness is controlled by spherical glass spacers and is typically about 5 $\mu$m. The experiments were conducted at a temperature of 20°C. The melting of LC nematic phase was performed by passing a small current via ITO covered glass slide. The experimental system has several response peaks in the 100-200 kHz range, with the strongest peak at 126 kHz, which was selected for the data acquisition. 

\textbf{The reconstruction of the director field, the optical phase retardance}. A custom-built fast multi-polarization video-microscopy setup was used to detect and quantify the internal order of the active liquid crystal with high spatial and temporal resolution. The schematics of the developed optical system is shown in Fig. S1a, see also Ref.\cite{katuri2022motility}. The LC film is illuminated with monochrome (590 nm) circularly polarized light. The camera sensor (Sony IMX250MZR), consisting of a conventional light-sensitive sensor overlaid by an array of polarizers, captures four polarized images at different orientations $0^0$, $45^0$, $90^0$, and $135^0$ simultaneously. 
An example of a raw image is shown in  Fig. S1(b). Each 2x2 pixels block (highlighted by the red square for illustration) contains information about the direction and the degree of polarization.  
Disodium cromoglycate (DSCG) liquid crystal in the nematic phase alters the polarization of light as a birefringent material. If DSCG is illuminated by the circularly polarized light and observed by the polarization camera, the local orientation of main optical axes, the optical retardance, and, correspondingly, the director field can be reconstructed from each 2x2 pixels block by using linear algebra (Jones calculus) and treating liquid crystal as a linear phase retarder. The schematic representation of such reconstruction is shown in Fig. S1(c).
For the reconstruction of the director field and the order parameter we first compute the relative variations of intensities $\widetilde{P_i}=P_i/P_0-1$, where $P_i$ is the pixel brightness in the same pixel block (see Fig. S1b,c), $P_0=1/4 \sum\limits_{i=1..4} P_i$ is the average brightness used for the binned image, Fig. S1(e). The local optical retardance $\Delta$ (proportional to the local order parameter) is calculated as:

\begin{equation}
\Delta=\frac{1}{2} \left( \sqrt{\widetilde{P_1}^2+\widetilde{P_2}^2}+\sqrt{\widetilde{P_3}^2+\widetilde{P_4}^2} \right) 
\label{de1ta}
\end{equation}

The local angle of the director field $\varphi$ can be computed by two equivalent formulas: 

\begin{equation}
\varphi_1=\frac{1}{2} \sign \left({\widetilde{P_1}} \right) \arccos \frac{\widetilde{P_2}}{\sqrt{\widetilde{P_1}^2+\widetilde{P_2}^2}}
\label{alpha1}
\end{equation}

\begin{equation}
\varphi_2=-\frac{1}{2} \sign \left({\widetilde{P_4}} \right) \arccos \frac{-\widetilde{P_3}}{\sqrt{\widetilde{P_3}^2+\widetilde{P_4}^2}}
\label{alpha2}
\end{equation}

For improved accuracy, we find the average of the two vector fields produced by doubled angles $\varphi_1$ and $\varphi_2$ with the amplitude proportional to the $\Delta$. The spatial Gaussian filter is applied for smoothing the data and reducing the noise. The local angle of the original director field is calculated as half of the angle of the averaged vector field. For the data presented in the main text the Gaussian filter with the $\sigma$= 2 pixels, which corresponds to $\approx$ $0.9$ $\mu m$, was applied. 

\textbf{Vorticity of the flows.}
Hydrodynamic flows that emerged in AELC were estimated by performing Particle Image Velocimetry (optical flow) on images obtained from the polarization camera with 2x2 digital binning. The optical disturbances caused by the non-uniform distribution of optical retardance in the LC film during active turbulence create dark and bright spots and stripes in LC media visible by conventional bright field microscopy. By tracking the motion of these spots we reconstruct the structure of hydrodynamic flows in AELC at different levels of activity.

\textbf{Activity patterns.}
The activity of AELC in the experiments is found in clusters with varying shapes and forms and not evenly spread throughout the droplet. The majority of these active clusters tend to be found near the LC/air interface, while isolated ones may be found in the bulk.
The clusters emerge due to the non-trivial propagation of acoustic waves in LC, which undergoes attenuation, scattering, reflection, and diffraction of topological defects\cite{mullen1972sound,pereira2013metric,miyano1979sound}. 

\begin{figure*}[h]
\centering
\includegraphics[width = 1\textwidth]{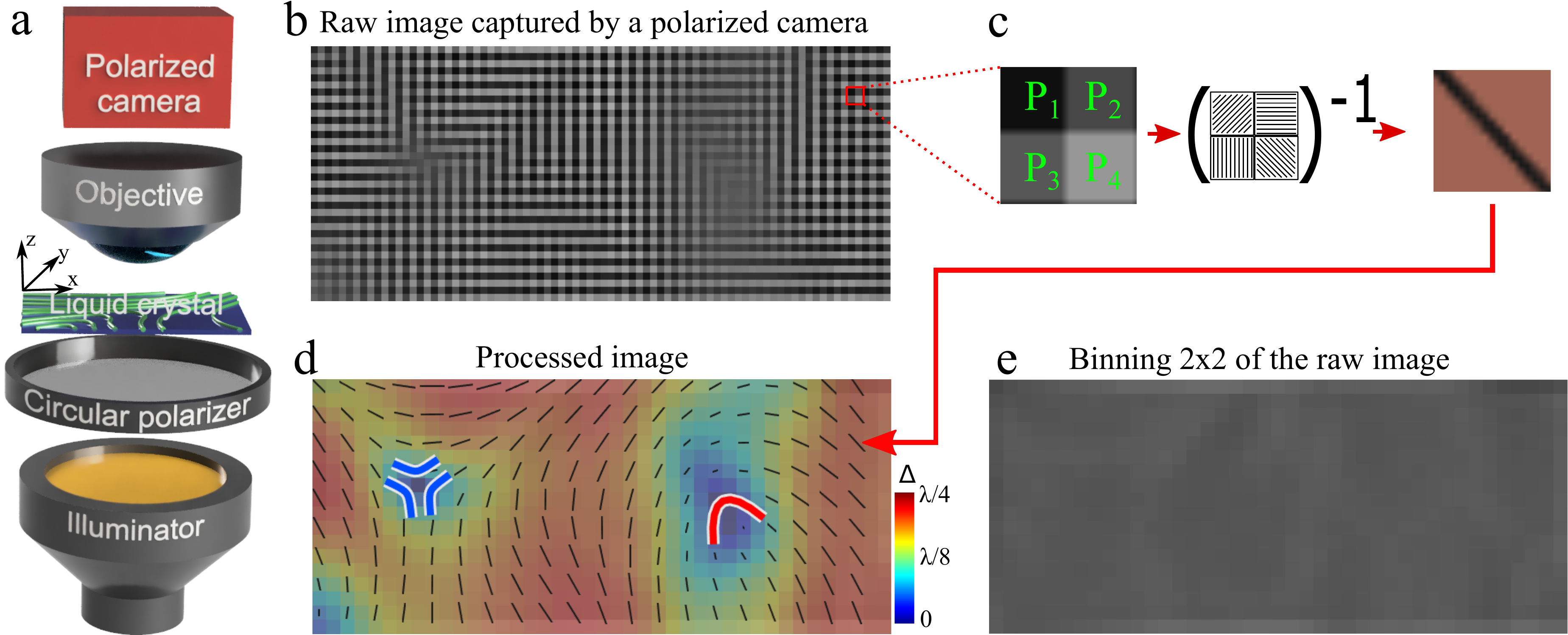}
\caption{\label{figS2} (a) Schematic representation of the optical system used in our experiments. The liquid crystal film is illuminated by the polarized monochrome light and observed with a polarized camera. (b) Example of the raw image captures with the polarized camera. (c) Information from each 2x2 pixel block is processed to reconstruct the local order parameter and the orientation of the director field. (d) An example of the reconstructed image obtained from (b). Short black lines illustrate the director field and the color depicts the phase retardance. (e) The image of the LC film captures by the non-polarized camera or with the binning 2x2 by the polarized camera. 
}
\end{figure*}

\bibliography{refs}
\bibliographystyle{rsc}

\subsection*{Movies description:}
Movie 1: Emergence of undulation instability in two distinct domains with different orientations of the director. Color illustrates the local orientation and short black lines depict the original local orientation before application of the acoustic wave.  

Movie 2 and 3: Advection of passive tracers by flows during the development of undulation instability

Movie 4: Development of undulation instability and transition to the turbulent motion. Topological positive and negative defects are marked by  red and blue symbols correspondingly. 

Movie 5: Hydrodynamic vortices in AELC.

Movie 6: Rotation of two positive topological defects in the core of a vortex.

Movie 7: Velocity field around a hydrodynamic vortex. 

Movie 8: Alternation of vortices rotation directions between different active domains. 

Movie 9: Switching the direction of rotation with the frequency change from 128kHz to 139kHz. The video is being played at a speed 3x faster than real-time.
